# Enhanced spatial resolution of Terahertz spectroscopy via semiconductor photoexcitation

Daniel Krotkov[1,2], Eli Flaxer[2,3], and Sharly Fleischer*[1,2]

[1]Raymond and Beverly Sackler Faculty of Exact Sciences, School of Chemistry, Tel Aviv University 6997801, Israel.
[2]Tel-Aviv University center for Light-Matter-Interaction, Tel Aviv 6997801, Israel
[3]AFEKA – Tel-Aviv Academic College of ENGINEERING, 69107 Tel-Aviv, Israel
*email address: sharlyf@tauex.tau.ac.il

**Abstract**

We utilize the photoexcitation of a semiconductor material as a 'reflectivity switch' for a broadband terahertz field. We show that judicious use of this switch enables temporal characterization of the THz field with spatial resolution significantly surpassing the diffraction limit of the terahertz and provides desirable means for spatio-temporal terahertz spectroscopy.

**Introduction**

Electromagnetic fields in the terahertz frequency range ($10^{11}$-$10^{13}$ Hz) have become broadly available thanks to the development of various table-top radiation sources starting from the Auston switch [1] and photoconductive antennas [2] , through optical rectification of short laser pulses in electro-optic crystals [3] (GaP, ZnTe, LiNbO3), organic crystals (DAST, DSTMS, OH1, HMB-TMS) [4–8], mixed-field THz generation in gas plasmas [9,10], metamaterials [11,12] and even in thin water films [13]. Within the plethora of research fields and prospected applications [14–19] enabled by generation and detection of THz fields, substantial efforts have been focused on the interaction between THz fields and semiconductor materials. Specifically, the ability to characterize the complex THz field provides access to the complex dielectric function and frequency dependent complex conductivity from which electronic properties such as the mobility and lifetime [20,21] are extracted [20,22]. The excitation of semiconductors by an optical pulse with above band-gap frequency invokes abrupt changes in its optical properties that manifest in an abrupt change in both THz transmission and reflection. Time-resolved optical pump – THz probe spectroscopy [21,23–25] (TROPTS) is used for monitoring such changes with sub-picosecond resolution [26].

While THz frequencies are 2-3 orders of magnitude lower than the visible frequencies, they remain beyond the range of direct detection via electronics and as such, are usually detected via electro-optic sampling in the far-field [3]. Their long wavelengths (hundreds of μm to few mm) invoke significant limitations on the spatial resolution of THz spectroscopy [27]. In addition, THz fields remain invisible due to the severe lack in imaging apparatuses (such as sensitive THz cameras, the development of which remains a central technological challenge of THz research).

Commercial THz cameras require intense THz pulses with peak fields ~1MV/cm [28,29], however lack any temporal resolution required for THz time-domain spectroscopy. We note however, that imaging with resolution beyond the diffraction limit was demonstrated using laser filament in air plasma [30] and via near field THz microscopy [31,32].

In this work we utilize ultrashort near-IR photoexcitation of a Silicon semiconductor wafer (SiSC) as a switchable THz reflector [24,26,33–37] to demonstrate super-resolution THz spectroscopy and for spatio-temporal characterization of THz fields at the surface of the sample. The paper is organized as follows: Section 1 describes the home-built THz reflection spectrometer used in this work. In section 2 we characterize the photo-induced THz reflectivity. Section 3 demonstrates super-resolution THz spectroscopy, and section 4 shows the spatially resolved THz field impinging at the sample surface.

1. **Experimental System**

The experimental system used in this work is a home-built Optical-pump THz-probe time-domain spectrometer (OPTP-TDS) in reflection configuration [38]. A near-IR pulse (800nm, 100fs duration) is split in three parts to form a THz generation beam, a strong pulse for optical excitation and a weak read-out pulse for the THz detection setup. **Beam 1** with ~0.7mJ/pulse generates single cycle THz pulses via tilted-pulse front optical rectification in a LiNbO$_3$ crystal [39,40]. Typical THz fields generated in our system cover the 0.1-1.2THz frequency band (figure 2b) and reach peak field amplitudes of ~60kV/cm. The THz beam is routed through a 4-f reflective optics setup consisting two pairs of off-axis parabolic reflectors (1$^{st}$ and 4$^{th}$ at 90°, 2$^{nd}$ and 3$^{rd}$ at 15°, see Fig.1). The THz beam is reflected from the sample/reference mounted on a home-built vacuum holder situated at the 2-f position. Vacuum is provided by a Venturi-pump in order to avoid mechanical vibrations that may be induced by scroll pumps. The vacuum holder enables fast and easy switching between the samples and reference without need for re-alignment of the reflected beam direction. **Beam 2** is used as a readout pulse for the characterization of the THz field at the output of the 4-f system via time resolved electro-optic (EO) sampling [3,41]. In this method a weak, 800nm readout pulse is routed through an EO crystal (GaP in our setup) situated at the focus of the last (4$^{th}$) off-axis parabolic mirror. The THz field induces birefringence in the crystal that is experienced by the weak readout pulse and alters its polarization state. The latter is directly proportional to the THz electric field, hence by scanning the delay between the THz and the weak readout pulse, one is able to characterize the complex electric field (amplitude and phase) as shown in Figure 2a. [42]. **Beam 3** is a 100fs, 800nm pulse used for photoexcitation of the SiSC sample at a computer controlled delay with respect to the THz field. The large f-number provided by 4-f THz setup (THz beam diameter ~20mm, focal length of ~381mm) results in a focused beam waist of ~10mm at the sample while the photoexcitation beam can be focused down to <50μm at the surface of the sample. In order to detect the changes in THz reflectivity at the overlap area of the two beams, we modulate beams 1 and 3 at two different frequencies using two optical choppers and lock-In detection of the signal at their difference frequency (see section 3). In this work, we detect the modulations in the THz reflection from a localized area with diameter $< 100 \mu m$ – i.e. significantly smaller than the diffraction limited THz spot defined by its the long wavelengths contained in the THz beam ($250 \mu m$ ~ 2mm).

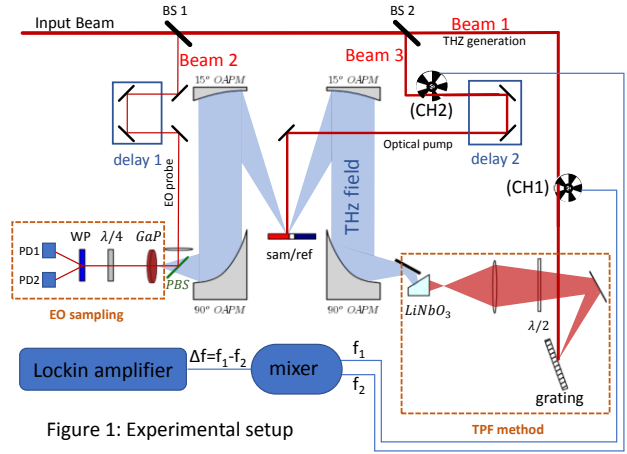

Figure 1: Experimental setup

## 2. Photo-induced THz reflectivity

In this section we explore the THz reflectivity ($R_{THz}$) induced upon photoexcitation of the SiSC sample. Figure 2 depicts the THz field that is regularly generated in our setup in both the time domain (Fig.2a, obtained by EO sampling) and frequency domain (Fig.2b, obtained by Fourier Transformation of the time domain shown in Fig. 2a).

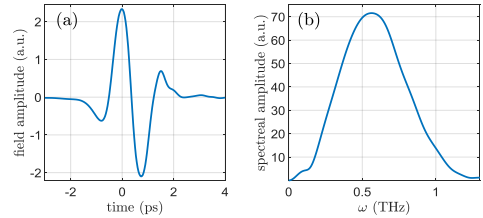

Figure 2: representative single-cycle THz pulse generated in our THz-TDS system. (a) time domain measured via EO sampling and (b) its frequency domain obtained by Fourier transformation of (a).

In order to obtain Fig.2 we positioned a gold-coated mirror at the sample position to fully reflect the input THz field ($R_{THz}^{mirror}$). When the gold mirror is replaced by a SiSC wafer, the reflected THz field amplitude is reduced by ~40% across the entire THz bandwidth as shown in figures 3 (time domain) and in Fig.7 (frequency domain). Note that intrinsic SiSC (undoped) are commonly used as beam splitters for the THz frequency range [43,44].

In order to explore the change in $R_{THz}^{SiSC}$ upon photoexcitation we apply beam 3 to photo-excite the SiSC wafer, as shown in Figure 3. The blue curve shows the time domain of the input THz pulse, reflected-off a gold mirror. The red curve depicts the case of an unexcited SiSC wafer, namely with beam 3 blocked. In this case, $\sim 57\%$ of the THz field amplitude is reflected from the front surface of the sample and detected by the EO sampling detector at t=0. The refracted THz field propagates through the sample, reflects from the back surface and reaches the EO sampling detector at $t \approx 14 ps$ (the time it takes the THz field to propagate back and forth through the $500 \mu m$ SiSC wafer). The $\pi$-phase shift of the back surface reflection relative to the front is readily observed by the EO sampling detection as expected upon refraction from high refractive index ($n_{0.2-1.5THz}^{SiSC} \approx 3.42$ [45]) to low refractive index ($n_{air}$~1). The green curve shows the reflection from a freshly excited SiSC, resulting in a dramatically enhanced $R_{THz}^{SiSC}$ from the front surface with close to perfect reflectivity (~93% of the mirror reflectivity). Correspondingly, the transmission through the front surface is severely reduced and the back surface reflection is even smaller as it goes through the excited front surface twice.

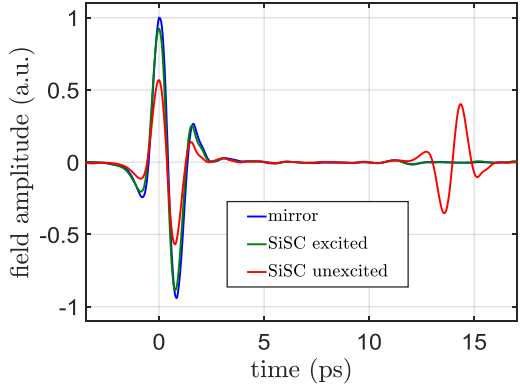

Figure 3: time-resolved THz field reflected from a 500μm thick SiSC wafer unexcited (red), photoexcited (green) and the reflection from a mirror (blue). Note the dramatic increase in the amplitude of the front surface reflection and decrease in that of the back surface reflection.

By varying the delay between the THz and optical pump and recording the full THz field via EO sampling we can obtain a time-resolved 2D map shown in the Supplementary information section SI.1.

**Characterization of the THz reflectivity switch**

The experimental results of Fig.3 provide ultrafast photo-switching of the THz reflectivity. For utilizing the switch in practical applications it is important to characterize the time scales of the photoexcitation and its dependence on the photoexcitation pulse intensity.

Figure 4 shows the time evolution of $R_{THz}^{SiSC}$. Here we park the EO readout pulse at the peak of the THz field (marked in inset (a)) and vary the arrival time of the optical excitation pulse. As can be seen by the abrupt change in $R_{THz}^{SiSC}$, the SiSC is in fact a one-way switch, namely it turns from low THz reflectance ('OFF') to high ('ON') over ~5 ps (inset (b)), and decays back to the OFF state in few nanoseconds [23] (hence appears as constant during the first 250ps spanned here). Note the dramatic change in $R_{THz}^{SiSC}$ upon photoexcitation as the amplitude of the reflected THz field increases by a factor of ~1.4 – from ~60% to ~88% in field amplitude. Once the SiSC switch is fully ON (~5ps past photoexcitation) it remains stable for several hundreds of picoseconds, i.e. much longer than the time it takes the reflection from the back side to traverse the semiconductor (~$14ps$ as shown in Fig.3)

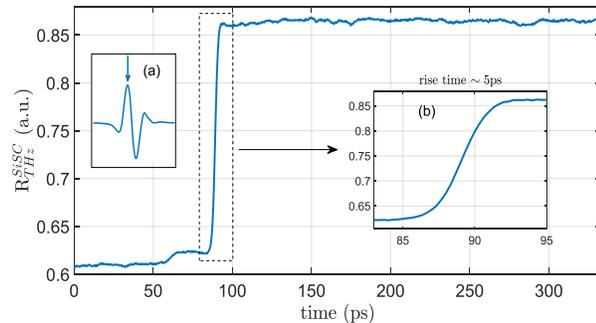

Figure 4: THz reflectivity from the front surface as a function of the THz-photoexcitation time delay. The readout pulse was parked at the peak of the THz field (marked by the arrow in inset (a)) and the photoexcitation timing was scanned across the entire delay stage. At t>95ps the photoexcitation precedes the THz arrival and results in increased reflectivity that remains stable, with no apparent decay for the timespan of the measurement (namely, at least for 250ps). Inset (b) is the enlarged region of temporal overlap between the excitation and THz field from which the switching time for THz reflectivity is estimated ~5ps.

## Dependence on optical pulse intensity

As the intensity of the photoexcitation pulse increases, the carrier density at the surface of the SiSC increases, elevating the surface conductivity and increasing $R_{THz}^{SiSC}$ correspondingly.

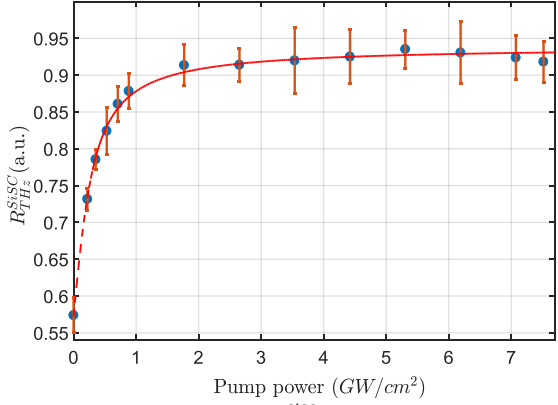

Figure 5: Dependence of $R_{THz}^{SiSC}$ on the photoexcitation beam power, in $GW/cm^2$ (per pulse). Each point represents the peak of a THz transient reflected off the SiSC wafer (as shown in Fig.4 inset (a)), normalized by the mirror reflectivity. The optical beam diameter was 12mm and the pulse duration 100fs. The trend line (red curve) is given by the arctan function, chosen as the best fitting standard saturating function.

Figure 5 shows the experimental $R_{THz}^{SiSC}$ for a photoexcitation preceding the THz field by ~10 ps. Naturally $R_{THz}^{SiSC}$ cannot surpass that of the perfect reflector ($R_{THz}^{mirror}$, aluminium or gold mirror), hence saturates at ~93% for optical pulse intensity of $> 2-3 \frac{GW}{cm^2}$. At higher intensities, the sample may experience laser induced damage and/or heating. The latter may manifest in an increased thermally-induced THz reflectivity and compromised signal level due to background thermal reflectivity.

All of the above experiments utilized a collimated optical beam with diameter $\sim 12mm$ as it emerges at the output of the laser amplifier. In what follows we reduce the diameter of the photoexcited area in order to induce a spatially localized THz reflectivity switching. However, as the optical beam diameter is reduced significantly below the THz beam diameter (the latter is ~10mm in our setup), the detection of the THz reflectivity modulation becomes highly demanding. In fact, with the optical beam diameter reduced to 1mm, we can hardly detect $\Delta R_{THz}^{SiSC}$. As demonstrated in the remaining of the paper, by incorporating a differential chopping detection scheme [46–48] the above task becomes feasible. Briefly, by modulating both the THz and photoexcitation beam (using CH1 and CH2 in Fig.1 respectively) at two different modulation frequencies and triggering the lock-in at their phase-locked difference frequency [49] we can selectively detect the change in THz reflectivity ($\Delta R_{THz}^{SiSC}$) emanating solely from the localized overlap area of the two beams. For detailed description of the differential chopping see section SI.2 in the supplementary information [47,48]).

## 3. Super resolved THz reflection spectroscopy

Figure 6 depicts the experimental signal obtained with differential chopping detection. The THz beam diameter at the surface of the SiSC was $\sim 10mm$ while the optical photoexcitation beam is focused by a lens ($f = 150mm$) to a spot size $< 100\mu m$ at the centre of the THz beam on the SiSC.

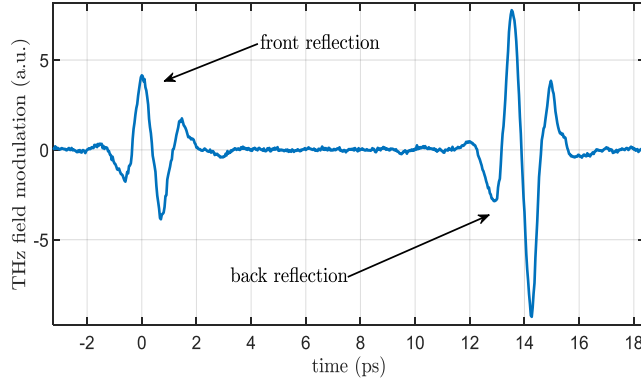

Figure 6: Measurement of the THz transient reflected from the photoexcited SiSC wafer with the optical pump focused to $\sim 100\mu m$ spot and attenuated to $\sim 0.5\mu J/pulse$. The signal is detected at the difference frequency of 500Hz (THz) and 333Hz (photoexcitation) and represents the photoinduced change in THz amplitude ($\Delta R_{THz}^{SiSC}$).

The advantage of the differential chopping detection is demonstrated in its full glory, enabling the probing of $\Delta R_{THz}^{SiSC}$ selectively from an area $\sim 4$ orders of magnitude smaller than the THz beam area on the sample surface. As will be further demonstrated, the overlap area from which we detect the THz switching responses is significantly smaller than the THz wavelength, thus experimentally realizing super-resolved THz detection. Note that unlike the results of Fig.3 where the amplitude of the THz reflected from the back SiSC surface is lower than that of the front surface (as expected), in Fig.6 we find that the modulation amplitude of the back surface reflection surpasses that of the front surface significantly. This is due to the differential chopping detection that unveils the change in THz amplitude $\Delta R_{THz}^{SiSC}$, excited vs. unexcited, rather than the THz amplitude directly. Therefore the back surface reflection that experiences the effect of the photoexcitation switching twice (upon entering and exiting the excited SiSC) instead of only once (for the front surface reflection), results in a larger $\Delta R_{THz}^{SiSC}$ back-surface signal than that of the front. The same holds for a non-focused beam, as shown in SI.3 in the supplemental information). Another indication for the $\Delta R_{THz}^{SiSC}$ measurement (Fig.6) instead of $R_{THz}^{SiSC}$ (Fig.3) is provided by comparing the phases of the front and back reflection signals: while for $R_{THz}^{SiSC}$ the back surface signal is out of phase with respect to the front surface signal (Fig.3), in $\Delta R_{THz}^{SiSC}$ they are in phase (Fig.6) (see SI.3 for details).

In Figure 7 we show and compare the spectral content of the front surface THz reflection for few scenarios (by fourier transformation of the detected time-domain signals).

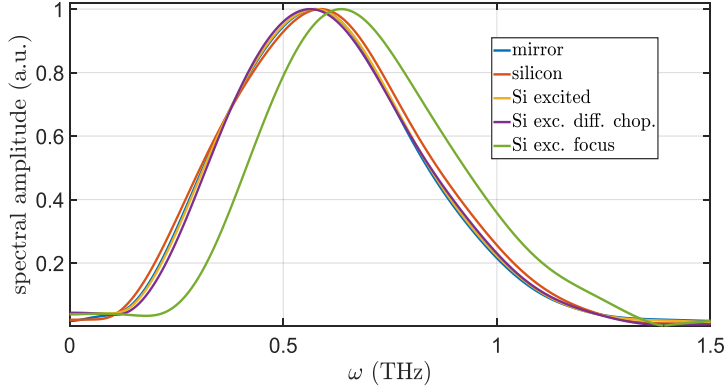

Figure 7: Normalized spectral amplitude of THz signals reflected from: (Blue) - Gold coated mirror, (Red) - Unexcited SiSC wafer, (Yellow) - Photoexcited SiSC wafer with ~12mm optical pump diameter, (Purple) - Same as yellow only measured with differential chopping detection. (Green) – SiSC wafer with a focused beam excitation (~100um diameter, front surface reflection of Fig.6).

The blue curve shows the spectral amplitude of the THz field reflected from the mirror. The orange and yellow curves show the spectral content of the THz field reflected from the unexcited (orange) and photoexcited (yellow) SiSC sample respectively, using the full size of the pump, ~12mm diameter, 300 uJ/pulse and using the standard lock-in detection scheme at the modulation frequency of the THz field (500Hz). The purple curve shows the spectral amplitude of the signals obtained with differential chopping (i.e. the spectral amplitude of $\Delta R_{THz}^{SiSC}$) for the exact same parameters of the yellow curve. The green curve depicts the spectral amplitude of the front surface reflection shown in Fig.6 (photoexcited with a ~100μm, ~450nJ/pulse). As readily observed, the THz reflected from the SiSC wafer, whether excited or not, remains similar to that of the mirror in both detection schemes, reassuring that the spectral response of the SiSC is flat across the frequency range of our THz pulse. However, upon focusing the optical beam down to $< 100 \mu m$, the low THz frequency components are somewhat attenuated as shown by the green curve in figure 7. The latter is qualitatively explained by comparing the THz beam waist for selected spectral components of the THz pulse: consider a perfectly focused THz field spanning the 0.1-1.2 THz region. The beam waists of the different frequency components included in the pulse vary significantly since their associated wavelengths spread between $3mm\ (0.1 THz) to\ 250 \mu m (1.2 THz)$ (see SI.4 in supplementary information). Therefore, the ratio between the overlap area (the optical photoexcitation area) and the THz beam waist strongly depends on the THz frequency (or wavelength), with a relatively weaker response for longer THz wavelengths as observed in Fig. 7.

In order to get a better estimate for the spatial resolution provided by the photoexcitation pump – THz probe technique, we fabricated a resolution target on a SiSC wafer shown in Figure 8a. We imprinted by photolithography several arrays of gold stripes with varying thickness and spacing apart, ranging from 1mm to 100μm.

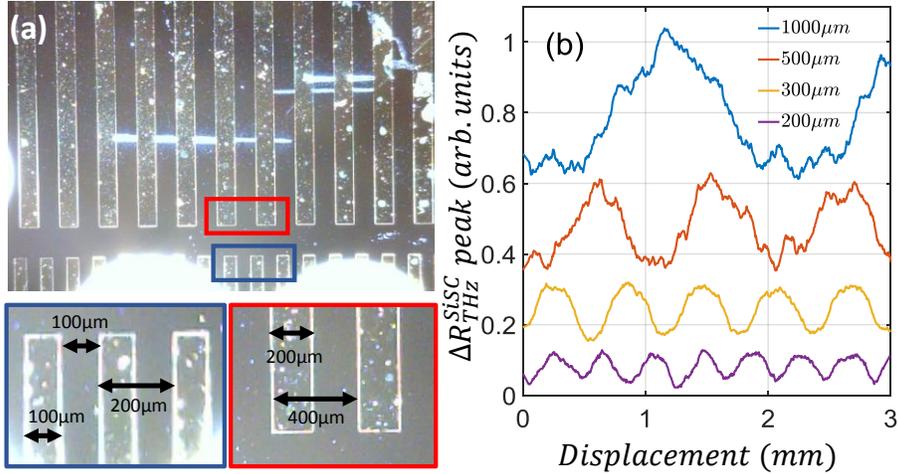

Figure 8: (a) microscope image of a small section of the resolution target with gold-deposited bars of 100μm and 200μm thickness on SiSC substrate. (b) Measurements of the peak of the THz signal modulation $\Delta R_{THz}^{SiSC}$ (similar to inset (a), figure 4) obtained by scanning the target across the interaction region of the optical and THz beams for several arrays of different thickness.

The experimental results of Fig.8b were obtained as follows: the THz field impinges at the surface of the gold-patterned SiSC sample of Fig.8a, covering several tens of the gold-deposited stripes. The optical pulse is focused to $\sim 100$μm at the centre of the THz area. We set the timing of the optical readout pulse at the peak of the THz field, as shown in inset (a) of Fig.4 and record the signal via the differential detection method. The sample was mounted on a computer-controlled stage and horizontally displaced across the beam overlap area for few millimetres (given by the x-axis of Fig.8). The resulting signal demonstrates the peak of $\Delta R_{THz}$ as a function of the position of the optical pulse on the wafer. Maximal modulation is obtained where the optical pulse is focused at the bare wafer surface while minimal (or no modulation) found when the optical pulse impinges at the gold stripes. THz modulation was observed down to an array spaced by 200μm. For smaller spacing we can hardly detect the modulation in $\Delta R_{THz}$ due to the large difference of the optical beam size and the THz beam size. In fact, the expected modulation is linear with the optical beam diameter and inversely proportional to the THz beam diameter. Both can be further optimized to achieve even higher spatial resolution. We note that observing the modulation at 200μm spacing between the gold features provides validation of our super-resolution THz method since the THz fields detected in Fig.8 (with THz bandwidth given by the green curve in Fig.7) contains a broad range of THz wavelengths, most of which are few-fold larger than the spatially resolved features of the sample. We note that a rough estimation of the optical beam diameter at the SiSC is provided by slightly increasing the optical beam power to carve a laser-induced damage on the surface of the sample. The width of this damaged line is $30 \sim 40$μm in diameter, namely $\sim 2 - 3$ fold smaller than the 100μm beam waist noted before and attributed to the nonlinear intensity dependence of the optical-induced damage [50] .

## 4. Spatial-temporal characterization of the THz focal area

The severe lack in THz imaging devices poses inherent difficulties to experimental THz research. For example, THz beam routing and THz focusing is typically performed with a visible laser source (such as He-Ne) that is replaced by the THz source at the last step of alignment. Spatial characterization of the THz focal area is performed by a pyroelectric sensor or bolometer that detects the integrated THz power transmitted through an aperture placed at the sample position. Hence, various possible deviations from the desirable Gaussian profile at the THz focus such as astigmatism, spatial chirp, etc. remain largely undetectable and unknown.

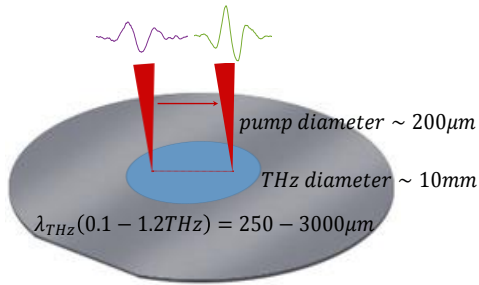

Figure 9, a sketch of the sweep scan of the optical pump across the THz beam. The pump probes localized time and phase information.

In what follows, we employ the optical-pump THz-probe for spatially-resolved temporal characterization of the THz field at the surface of the sample. We position the bare (unpatterned) SiSC wafer at the sample position and scan the position of the optical pulse timed to photo-excite the SiSC few tens of picoseconds before the arrival of the THz field. We scan the photoexcitation pulse with diameter $\sim 200 - 300\mu m$ across the THz focal area on the SiSC. Note that unlike the previous section, where the sample was translated across the overlap area of the optical and THz fields, here the sample position and the THz focal area are fixed while the optical pulse is scanned across as pictorially shown in Fig. 9.

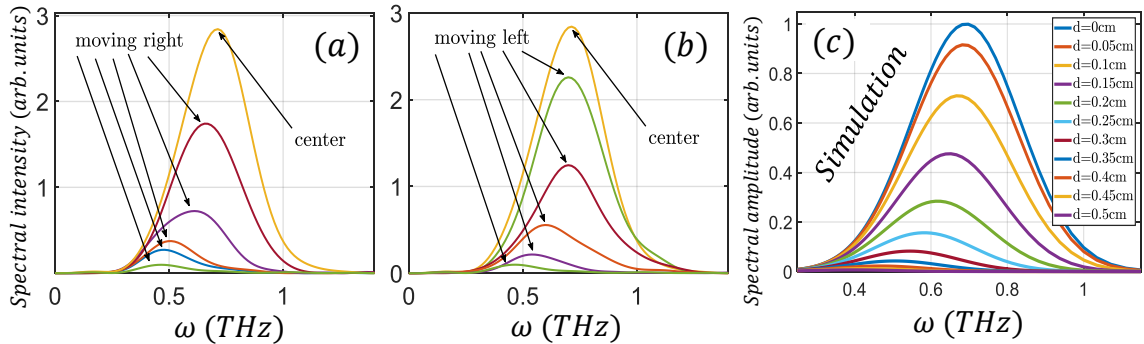

Figure 10, absolute value squared of the frequency domain of the horizontal scan of the THz beam at its focal point. The optical pump scanned the THz beam starting at the midpoint, taking 1mm steps to the right (a) and to the left (b). A Gaussian beam optics simulation is presented in (c) to support the results of (a) and (b), where d is the shift from the center.

For each position of the optical pulse we record a local THz transient. Figure 10 depicts the THz spectra obtained by Fourier transformation of the THz transients for a progression of the optical pulse position starting from the centre of the THz beam (yellow spectrum in Figs. 10a,b) and gradually moving to the right (Fig.10a) and left hand side (Fig.10b). The spectra in Fig.10a,b show that the local spectral content of the THz field shifts toward lower THz frequencies with increased distance from the centre of the THz beam (in both right and left directions). Fig.10c shows our simulated results where we took the THz frequency content of our initial experimental pulse (yellow in fig. 10a, b), calculated the beam waist as a function of frequency using Gaussian beam optics, and shifted away from the center. Clearly the higher THz frequencies (shorter wavelengths) have smaller beam waists than those of the lower THz frequencies (see SI.4), which results in stronger attenuation of the high THz frequencies as the optical beam is successively shifted away from the centre point. As readily seen, the simulation results of Fig.10c are in agreement with the experimental scans of Fig.10a and b, clearly demonstrating the same trend in the observed spectral shift.

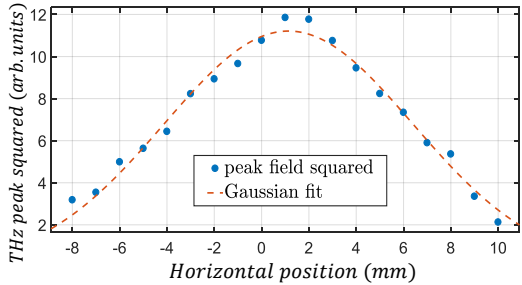

Figure 11: Intensity of the reflected THz field as a function of position across the beam fitted to a Gaussian beam profile.

Characterization of the spatially-resolved THz spectrum across the beam area requires recording of the full THz transient at each position of the optical beam. A downgraded, yet very useful 1D version for probing the spatial intensity profile of the THz beam is demonstrated in Fig. 11. Here we parked the optical readout pulse at the peak of the THz field (as shown in inset (a) of Fig. 4a) and scanned the horizontal position of the photoexcitation pulse. For each position of the optical pulse we record only the peak of the THz field. Squaring the latter and plotting against the position of the photoexcitation pulse reveals the Gaussian intensity profile of the THz beam (dashed red curve). Mapping the spatial THz intensity profile can be generalized to 2D by "raster-scanning" of the optical pulse position in both the horizontal and vertical directions.

**To conclude**, in this work we utilized the optical photoexcitation of a Silicon wafer as an ultrafast THz reflectivity switch. Using THz-TDS in reflection configuration we characterized the time-scales of the switch and its dependence on the intensity of the photoexcitation pulse. Implementing the differential chopping detection technique significantly enhances the detection sensitivity and allowed monitoring the change in THz reflectivity with dramatically improved spatial resolution. Obtaining THz spectra from a spot size several folds smaller than the THz wavelength is a manifestation of super-resolution THz spectroscopy in the far field. The demonstrated ability to extract THz spectra from a localized area of interest (defined by the area and position of the photoexcitation pulse) provides highly desirable means for spatial and temporal characterization of the THz field at a position of interest. We note that the capabilities demonstrated and discussed in this work are merely few examples for desirable applications in THz science and technology – where spatial imaging devices are severely lacking.

The authors acknowledge support of the Wolfson Foundation (Grants No. PR/ec/20419 and PR/eh/21797), the Israel Science Foundation – ISF (Grants No. 1065/14, 926/18, and 2797/11 by INREP—Israel National Research Center for Electrochemical Propulsion.

The authors declare no conflicts of interest.